\documentclass[useAMS,usenatbib,referee]{mn2e}
\usepackage{graphicx}

\title[Radio emission from the colliding winds of $\eta$ Carinae]
{Wind-wind collision in the $\eta$ Carinae binary system II: Constrains to the binary orbital parameters from radio %%@
emission near periastron passage}
\author[ Z. Abraham, D. Falceta-Gon\c{c}alves, T. P. Dominici, A. Caproni and V. Jatenco-Pereira]
{Z. Abraham$^{1}$\thanks{E-mail:zulema@astro.iag.usp.br}, D.Falceta-Gon\c{c}alves$^{1}$, T. Dominici$^{1}$, A. %%@
Caproni$^{1}$, V. Jatenco-Pereira$^{1}$ \\
$^{1}$Instituto de Astronomia, Geof\'\i sica e Ci\^encias Atmosf\'ericas, Universidade de S\~ao Paulo, PO Box 3386, %%@
01060-970, S\~ao Paulo, Brazil}

\begin{document}

\date{ }

\pagerange{\pageref{firstpage}--\pageref{lastpage}} \pubyear{2005}

\maketitle

\label{firstpage}

\begin{abstract}
In this paper we use the 7 mm and 1.3 mm light curves obtained during the 2003.5 low excitation  phase of the $\eta$ %%@
Carinae  system to constrain the possible parameters of the binary orbit. To do that we assumed that the mm wave emission %%@
is produced in a dense disk surrounding the binary system; during the low excitation phase, which occurs close to %%@
periastron, the number of ionizing photons decreases, producing the dip in the radio emission. On the other  hand, due to %%@
the large eccentricity, the density of the shock region  at periastron is very high and the plasma is optically thick for %%@
free-free radiation at 7 mm, explaining the sharp peak that was observed at this frequency and lasted for about 10 days.  
From the shape and duration of the peak  we were able to determine the orbital parameters of the binary system, %%@
independently of the stellar parameters, such as mass loss rates, wind velocities or temperature at the post-shock region.
\end{abstract}

\begin{keywords}
stars: individual ($\eta$ Car) 
binaries: general
stars: variable
radio continuum: general
\end{keywords}
      
\section{Introduction}
The occurrence of the predicted 2003.5 low excitation event in the $\eta$ Carinae system, extending from radio to X-rays %%@
left little doubt about its periodic behavior  (Laj\'us et al. 2003, van Genderen \& Sterken 2003, Smith et al. 2004, %%@
Whitelock et al. 2004, Abraham et al. 2005, Corcoran 2005, Stahl et al. 2005,  Weis et al. 2005).
Although the binary nature of the system seems to explain the strict periodicity as well as the  high intensity of the %%@
X-ray emission, attributed to wind-wind collision (Pittard et al. 1998, Ishibashi et al. 1999), orbital parameters, like %%@
eccentricity, epoch of periastron passage and orientation of the orbit relative to the observer are not well defined %%@
(Davidson 1997, van Genderen et al. 1999, Pittard \& Corcoran 2002). 
Besides, other observed properties like the optical and UV light curves and the recently discovered He\,{\sc %%@
ii}\,$\lambda$\,4686\,\AA \, emission (Martin et al. 2004, Steiner \& Damineli 2004) seem to be more appropriately %%@
described by periodic shell events.    

Falceta-Gon\c calves, Jatenco-Pereira \& Abraham (2005, Paper I) were able to conciliate the two hypothesis, still assuming %%@
a binary system to explain the strong X-ray emission, but  also taking into account that, near periastron and because of %%@
the highly eccentric orbit, the wind behind the shock cools rapidly and forms dust grains, which absorb part of the optical %%@
and UV flux, as in a shell-ejection event. For this process to be effective, at periastron the secondary star must be %%@
located between $\eta$ Carinae and the observer, solving also the discrepancy between the orbital parameters derived from %%@
ground and space based observations of the radial velocities of the H recombination lines (Davidson 1997). 

Although radio waves are not affected by dust absorption, the radio light curves show also  periodic dips, similar to those %%@
found at X-rays (Cox et al. 1995, Abraham \& Damineli 1999, Duncan, White \& Lim  1997, Abraham et al. 1999). Based on high %%@
resolution radio observations, Duncan \& White (2003) suggested that the radio emission is produced by the free-free %%@
process in an extended ($2^{\prime\prime}$ radius) disk surrounding the  $\eta$ Carinae system;  the dips would be the %%@
consequence of a sudden decrease in the available ionizing UV flux, as expected in a shell-like event.

To better understand the mm-wave light curve during the 2003.5 event, Abraham et al. (2005)  made daily observations at 7 %%@
mm with the Brazilian Itapetinga radiotelescope and weekly observations at 1.3 mm with the ESO (European Southern %%@
Observatory) SEST radiotelescope at La Silla, Chile. The closely spaced observations provided not only the light curves %%@
with great precision, but also showed an unexpected increase in flux density, which could not be explained by free-free %%@
wind emission from the individual stars. Using reasonable values for the physical conditions at the shock during periastron %%@
passage,  they showed that the observed excess flux density is compatible with optically thick free-free emission produced %%@
at the wind collision site.

In this paper we show that the  mm-wave light curve during periastron passage, after subtracting the disk contribution, can %%@
be explained by simple geometric  arguments, if the free-free emission originates in the conic surface defined by the %%@
wind-wind collision interface.  Furthermore, we show that the orbital parameters are strongly bound by the light curve, and %%@
that periastron should occur when the secondary star is between $\eta$ Carinae and the observer, as also required to %%@
explain the X-ray light curve and shell-like events (Paper I).

In Section 2 we describe the emission model and the geometry of the shock surface. In Section 3 we present the results %%@
including expressions for the emission at different positions in the secondary's orbit and the determination of the orbital %%@
parameters. In Section 4 we discuss our results and compare them with those commonly accepted for the $\eta$ Carinae system %%@
and in Section 5 we summarize our conclusion. 

\section{Model for the mm-wave emission}

The observed light curves, obtained with the Itapetinga radiotelescope at 7 mm and with SEST at 1.3 mm are presented in %%@
Figure 1, where we can see the decrease in flux density superposed to a sharp peak at 7 mm, which reaches its maximum %%@
intensity on 29 June 2003. Details of the observations are found in Abraham et al. (2005).

We will assume that the observed free-free emission  is  formed by the contribution of both an extended ionized disk, which %%@
surrounds  the binary system, and  the  wind-wind shock material. At the beginning of the low excitation phase there is a %%@
sudden decrease in the  UV ionizing flux, as proposed by Duncan \& White (2003); the central denser part of the disk %%@
recombines in timescales of a few days and the flux density decreases. Also, since the recombination timescale  for the %%@
disk material depends on the unknown density distribution and we are interested mainly in the shock emission, for which the %%@
physical conditions are constrained by the orbital parameters, we will model the disk emission by an exponential decay %%@
function of the form:

\begin{equation}
S_{\rm disk}({\rm 7 mm})=a \exp(bt+c) + mt +n
\end{equation}  

Abraham et al. (2005) determined the free parameters $a$, $b$, $c$, $m$ and $n$ by fitting the data at both sides of the %%@
peak in the 7 mm light curve.  They also calculated  the physical conditions at the shock site required to obtain the %%@
observed peak flux density of about 1 Jy at this wavelength. The most favorable situation involved a high temperature %%@
optically thick plasma, as expected to be produced during periastron passage. 
In this paper we will obtain  simultaneously the orbital and disk emission parameters that reproduce the total observed %%@
light curves, both at 1.3 and 7 mm.

   \begin{figure}
      {\includegraphics{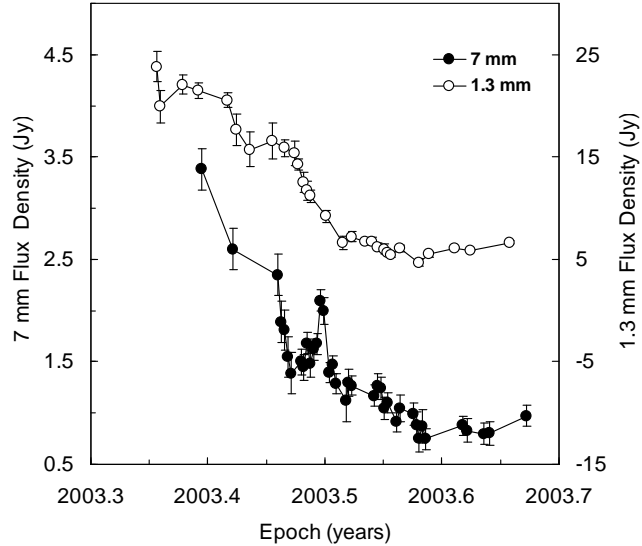}}
      \caption 
	  {mm-wave light curves of $\eta$ Carinae at the epoch of the predicted low excitation phase. Open circles correspond %%@
the 1.3 mm
	   emission (right axis), full circles correspond to 7 mm emission (left axis).}
         \label{figure1}
   \end{figure}

   \begin{figure}
   \includegraphics[bb=100 490 385 710,width=8cm]{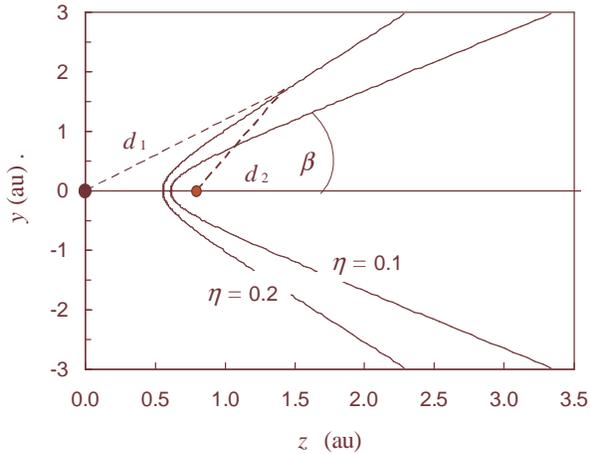}
      \caption{Surface of momentum balance between too colliding winds in a binary system. The stars are indicated by dots. %%@
The
	   separation between them corresponds to the $\eta$ Carinae system at periastron passage for an eccentricity $e = %%@
0.95$. The
 aperture angles of the asymptotic cone are 56\degr  and $44\degr $for $\eta$ values of 0.2 and 0.1, respectively. }
         \label{figure 2}
   \end{figure}

      \begin{figure*}
    \centering
   \includegraphics[bb=85 460 550 690,width=14cm]{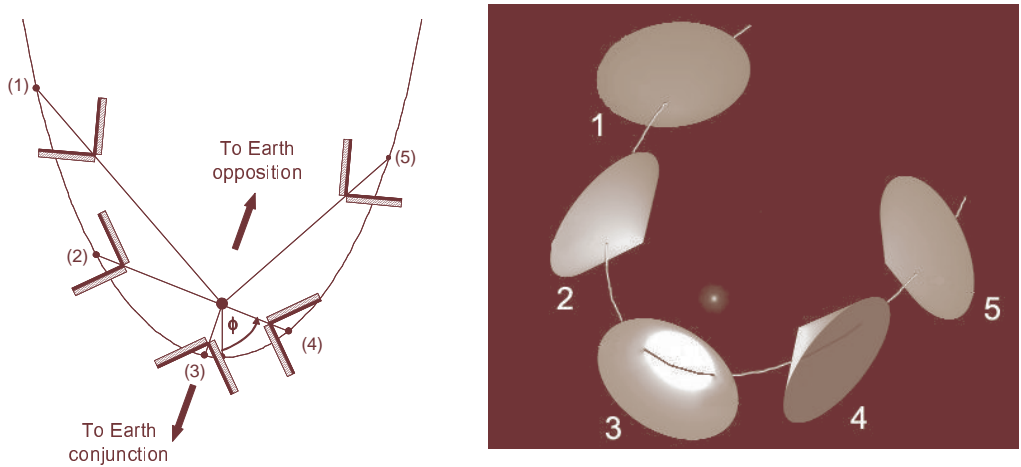}
%     {\includegraphics{Fig5.eps}}
      \caption 
{Left: schematic view of the binary orbit near periastron. The shock region is shown as a projected cone: the secondary and  %%@
$\eta$
 Carinae shocks are represented by the dark and the hatched regions, respectively. Points (2) and (4) represent the %%@
positions in
  the orbit for which the line joining the stars is perpendicular to the line of sight. The position of the observer at
   conjunction or opposition is represented by point (3). The angle $\phi$ represents the position
    angle of the secondary in its orbit, with $\eta$ Carinae at the focus of the ellipse. Periastron passage corresponds to %%@
$\phi
	 = 0\degr$. Right: artist view of the emitting surface with $\Phi = 0$.}        
 \label{figure3}
   \end{figure*}

\subsection{The free-free shock emission}

At mm wavelengths, the flux density  due to free-free emission must be calculated taking into account radiation transfer %%@
since, at the densities and temperatures found in the shock, the optical depth can be large. 
Using the Rayleigh-Jeans approximation we have:

\begin{equation}
dS(\lambda)=\frac{2kT}{\lambda^2}(1-e^{-\tau (\lambda)})\, d\Omega,
\end{equation}

\noindent
where $d\Omega$ is the element of solid angle subtended by the emitting region and $\tau (\lambda)$ the optical depth, %%@
calculated from:

\begin{equation}
\tau (\lambda)= \int_0^L \kappa_{ff}(\lambda)\, ds,
\end{equation}

\noindent
where $L$ is the depth of the emitting  region and $\kappa_{ff}(\lambda)$ is the free-free absorption coefficient, which at %%@
radio wavelengths can be calculated from:

\begin{equation}
\kappa_{ff} (\lambda)=\frac{3.7\times 10^8 h}{c^2}\,\frac{ n_i n_e g_{ff}(\lambda)\lambda^2}{T^{3/2}},
\end{equation}

\noindent
and $g_{ff}(\lambda)$ is the Gaunt factor for free-free emission. For the physical conditions at the shock, $T>10^5 Z^2$ %%@
and $hc/\lambda \ll  kT$, $g_{ff}(\lambda)$ can be approximated by:

\begin{equation}
g_{ff}(\lambda) = \frac{\sqrt{3}}{\pi} \ln \biggl(2.2\frac{kT\lambda}{hc}\biggr).
\end{equation}

The  optical depths at 1.3 and 7 mm can be determined from  the ratio of their respective flux densities. Although the %%@
sharp peak in the light curve was not observed at 1.3 mm, from Figure 1 we  can estimate  the contribution  of the shock %%@
emission at this wavelength. The emission from the extended disk is about 5 Jy at the minimum and the total 1.3 mm flux %%@
densities before and after the 7 mm maximum are about 11 Jy and 9 Jy, respectively. The difference gives the approximate %%@
value of 5 Jy for the expected shock contribution.  Assuming a flux density of 1 Jy for the 7 mm  peak emission, we find %%@
$\tau(7\; {\rm mm})\sim 4$ and  $\tau(1.3\; {\rm mm})\sim 0.2$. Therefore, close to the peak,  the 7 mm emission is %%@
optically thick and  the form of the light curve is completely  determined by the variation of solid angle subtended by the %%@
shock surface at each position of the binary orbit.

\subsection{The geometry of the shock surface}

The general form of the momentum balance surface between the two colliding winds was given by Stevens, Blondin \& Pollock %%@
(1992) as:

\begin{equation}
\medskip
\frac{dy}{dz}= \frac {(\eta^{-1/2}{d_2}^2+ {d_1}^2)y}{\eta^{-1/2}{d_2}^2z+{d_1}^2(z-D)},
\medskip
\end{equation}

\noindent
where $D$ is the distance between the  stars; $d_1$ and $d_2$ are  the distances of the primary and secondary star to the %%@
contact surface, respectively; the coordinates $z$ and $y$ are measured along the line that joins the stars and %%@
perpendicular to it, respectively; $\eta=\dot{M_s} v_s/\dot{M_p} v_p$, where  $\dot{M_p}$ and $\dot{M_s}$ are the mass
 loss rates of $\eta$ Carinae and the companion star, $v_p$ and $v_s$ their respective wind velocities.

The surface can be described asymptotically by a cone with an opening  angle $\beta$ that depends on $\eta$. Figure 2 shows %%@
a schematic view of the contact surface for two values of $\eta$ and a separation between the two stars equivalent of what %%@
is expected at periastron in the $\eta$ Carinae binary system.
The asymptotic angle defined by the surface is $56\degr $ for $\eta = 0.2$ and $44\degr$ for $\eta = 0.1$. 

When radiative losses are high, a thin and dense shock layer will form at both sides and very close to the contact  surface %%@
(Eichler \& Usov 1993).
Due to the orbital motion, the cone will be distorted beyond a distance from the cone origin $r > (P/2)v_p$ , where $P$ is %%@
the orbital period. For the $\eta$ Carinae system, with a period of 5.5 years and wind velocity of 700~km~s$^{-1}$, $r %%@
\approx 6 \times 10^{15}$ cm. 

In Paper I we assumed $\dot M_p = 2.5\times 10^{-3}$ M$_\odot$ yr$^{-1}$, $v_p = 700$ km s$^{-1}$ for $\eta$ Carinae and %%@
$\dot M_s = 10^{-4}$ M$_\odot$ yr$^{-1}$, $v_s = 3000$ km s$^{-1}$ for the secondary star, which results in $\eta = 0.2$. %%@
We  calculated the temperature and density at both the primary and secondary shocks near periastron as $T_p = 6\times 10^6$ %%@
K, $n_p = 2\times 10^{11}$ cm$^{-3}$ and $T_s = 2\times 10^8$ K, $n_p = 5.6\times 10^{9}$ cm$^{-3}$, respectively. Also, we %%@
pointed out that the actual temperature can be lower and, if we assume isobaric equilibrium, the density higher. Using %%@
these values, Abraham et al. (2005) estimated a radius for  the  projected  surface necessary to reproduce the 7 mm peak %%@
emission of about $10^{14} - 10^{15}$ cm. Under these circumstances, the shock surface can be approximated by the non %%@
distorted asymptotic cone. 

In Figure 3 we present a schematic view of the contact surface position and orientation with respect to the observer at %%@
different orbital phases, scaled for eccentricity $e=0.9$. 
$\phi$ is the position angle of the secondary star orbit relative to $\eta$ Carinae, with $\phi = 0$ at periastron and %%@
$\phi = \Phi$  at conjunction.
The dark and hatched regions represent the secondary and primary shocks, respectively. As we can see, if periastron occurs %%@
close to conjunction, for $\;\mid \phi + \Phi \mid\; <  \beta$ only the secondary shock is visible to the observer, for %%@
$90^\circ  >\;\mid \phi + \Phi \mid\; >  \beta $ only part of the emission from the secondary shock is directly visible, %%@
the remaining part is partially absorbed by the primary shock, which  also contributes with its own emission. For  all %%@
other position angles, only the primary shock surface is directly observed.
If periastron occurs close to opposition, the situation is reversed with respect to the secondary and primary emission.

\section{Results}
\subsection{Shock emission along the orbital phase}

%\subsection{Reproducing the observed 1.3 mm and 7 mm light curves}

For orbital inclination $90^\circ  \geq\;i \; >  \beta $, the observed emission is produced in the internal and external %%@
parts of the contact cone, which is located at a distance $D$ from the observer, and has an extension $R(\phi)$.
The optical depth of the primary and secondary shock regions are not known, except for the fact the maximum in the 7 mm %%@
light curve requires $\tau _{\rm s}\approx 4$ (Abraham et al. 2005). We will assume $\tau_{\rm s,p} \propto [d_{\rm %%@
s,p}(\phi)/d_{\rm s,p}(0)]^{-\alpha}$, where $d_{\rm s,p}(\phi)$ represents the distance from the shock to the respective %%@
star, with $\tau _{\rm s}(0)= 4$;  $\alpha$ and $\tau _{\rm p}(0)$ are free parameters to be determined from the fitting to %%@
the observed light curve.
We will also assume that the optical depths  are constant along the solid angle  subtended by the emitting surface (in fact %%@
$\tau _{\rm s,p}(\phi)$ is the average  over the solid angle). We will also assume that $R(\phi)$ varies as $[d_{\rm %%@
s,p}(\phi)/d_{\rm s,p}(0)]^2$ as estimated by Eichler \& Usov (1993) and $i=90^\circ$, as seems to be implied by the high %%@
resolution radio observations of Duncan, White \& Lim (1997). Smaller inclination angles, in the range $90^\circ  \geq\;i %%@
\; >  \beta $ will affect the projected surface area, requiring a larger value of $R$, which is anyway a free parameter in %%@
our model.
We  present the expressions for the total emission as a function of the orbital phase ($-\pi \leq \phi \leq \pi$) for the %%@
case  in  which periastron is close to conjunction, for the opposite case the role of the secondary and primary shock %%@
emission should be exchanged. We  include expressions for the emission from the primary shock, although its contribution is %%@
negligible compared to that of the secondary shock when the latter arises from an optically thick region, since $T_{\rm s} %%@
\gg T_{\rm p}$. 

\medskip 
\noindent
For $\;\mid \phi + \Phi \mid\; \leq  \beta $: 

\begin{equation}
S(\phi)= \frac{2kT_{\rm s}}{\lambda^2}\bigl[1-e^{-\tau_{\rm s} (\phi)}\bigr]\Omega _1(\phi),
\end{equation}
\noindent
with
\begin{equation}
\Omega _{1}(\phi) = \pi ab(\phi)
\end{equation}
\noindent
where $a=(R/D)\sin \beta$ and $b(\phi)=(R/D)\sin \beta\mid\cos(\phi+\Phi)\mid$.

\medskip
\noindent
For  $90^\circ  \geq\;\mid \phi + \Phi \mid\; >  \beta $ 
\begin{equation}
S(\phi) = S_{\rm 1s}(\phi) + S_{\rm 1p}(\phi), 
\end{equation}
\noindent
with
\begin{equation}
S_{\rm 1s}(\phi)= \frac{2kT_{\rm s}}{\lambda^2}\bigl[1-e^{-\tau_{\rm s} (\phi)}\bigr]\bigl[\Omega _{2}(\phi)+ 
        e^{-\tau _{\rm p}(\phi)}\Omega _{3}(\phi)\bigr],
\end{equation}
\noindent
and
\begin{equation}
S_{\rm 1p}(\phi)= \frac{2kT_{\rm p}}{\lambda^2}\bigl[1-e^{-\tau_{\rm p} (\phi)}\bigr]\Omega _{3}(\phi),
\end{equation}

\noindent
where

\begin{equation}
\Omega _2(\phi) =  ab(\phi)[\;\pi/2+(1-\zeta^2)^{1/2}+\arcsin \zeta\;],
\end{equation}

\noindent
and
\begin{equation}
\Omega _3(\phi) =  ab(\phi)\sin \beta \cos\beta \;(1-\zeta^2)^{1/2}\;
[\;\mid\sin(\phi+\Phi)\mid-\zeta^2\;],
\end{equation}

\noindent
with $\zeta=\;\mid \tan \beta /\tan (\phi+\Phi)\mid$.

\medskip
\noindent
For $180^\circ -\beta \geq  \;\mid \phi + \Phi \mid\; > 90^\circ $:
\begin{equation}
S(\phi) = S_{\rm 2s}(\phi) + S_{\rm 2p}(\phi), 
\end{equation}

\noindent
where
\begin{equation}
S_{\rm 2s}(\phi)= \frac{2kT_{\rm s}}{\lambda^2}e^{-\tau _{\rm p}(\phi)}\bigl[1-e^{-\tau_{\rm s} (\phi)}\bigr]\bigl[\Omega %%@
_{2}(\phi)+ 
        \Omega _{3}(\phi)\bigr],
\end{equation}
\noindent
and
\begin{equation}
S_{\rm 2p}(\phi)= \frac{2kT_{\rm p}}{\lambda^2}\bigl[1-e^{-\tau_{\rm p} (\phi)}\bigr]\bigl[\Omega _{2}(\phi)+
\Omega _{3}(\phi)\bigr],
\end{equation}

\medskip
\noindent
For $\;\mid \phi + \Phi \mid > 180^\circ- \beta  $: 
\begin{equation}
S(\phi) = S_{\rm 3s}(\phi) + S_{\rm 3p}(\phi), 
\end{equation}

\noindent
where
\begin{equation}
S_{\rm 3s}(\phi)= \frac{2kT_{\rm s}}{\lambda^2}e^{-\tau _{\rm p}(\phi)}\bigl[1-e^{-\tau_{\rm s}
 (\phi)}\bigr]\Omega (\phi),
\end{equation}
\noindent
and
\begin{equation}
S_{\rm 3p}(\phi)= \frac{2kT_{\rm p}}{\lambda^2}\bigl[1-e^{-\tau_{\rm p} (\phi)}\bigr]\Omega(\phi)
\end{equation}

   \begin{figure}
   \centering
   \includegraphics[bb=98 235 384 710,width=8cm]{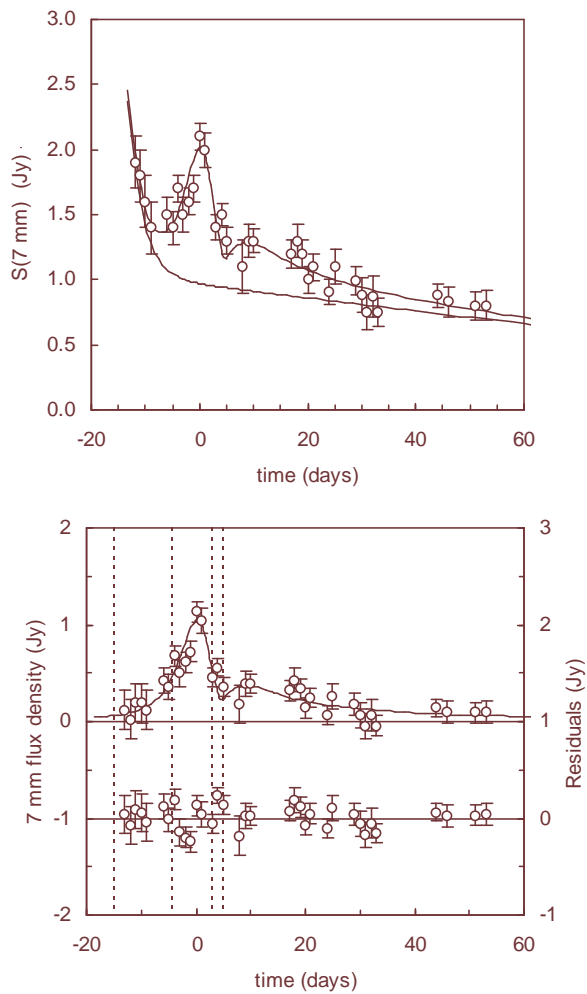}
      \caption 
	  {Model of the 7 mm light curve during the low excitation phase. Upper panel: fitted model together with the observed %%@
7 mm flux density (circles); the disk model emission is also shown. Lower panel: shock emission model (continuous line) and %%@
the observed shock contribution (circles), calculated as  the difference between the total observed flux density and the %%@
disk model  contribution. The residuals of the fitting, calculated as the difference between the observation and the total %%@
model emission (disk plus shock) are also shown. Vertical lines represents orbital position angles from left to right: %%@
$\phi + \Phi = -90^\circ, -\beta, \beta$ and $ 90^\circ$. Zero in the time axis corresponds to June 29, 2003 (JD = %%@
2452819).}
         \label{figure4}
   \end{figure}

\subsection{Determination of the orbital parameters from the mm-wave light curves}

We fitted the observed 7 mm light curve simultaneously to the shock emission model described in the previous subsection, %%@
using the parameters listed in the first column of Table 1, and  to the surrounding disk  emission, described at the %%@
beginning of Section 2. We found that each parameter affects differently the form of the light curve, in such a way that %%@
they do not let too much margin for variation. 

The width and shape of the light curve near the maximum  were affected mainly by the orbital parameters: the eccentricity %%@
$e$ determines how fast the secondary's position angle changes with time, the phase of conjunction $\Phi$ determines the %%@
asymmetry in the shape of the peak and the epoch of periastron passage defines its position in the light curve. 

The aperture angle of the emitting cone $\beta$ together with the rate at which the opacity changes, given by the parameter %%@
$\alpha$, were responsible for the  shape of the light curve at larger phase angles. For orbital angles $\;\mid \phi + \Phi %%@
\mid\; >  \beta $, the form of the light curve is also affected by the absorption and emission of the primary shock; if %%@
these last effects were important we should see a discontinuity in the light curve at the epoch in which  $\;\mid \phi + %%@
\Phi \mid\; =  \beta $, unless the  absorption of the secondary shock emission by the primary is exactly compensated by the %%@
emission from the primary shock. Since we did not see any discontinuity and we know that the emission from the primary %%@
shock is much smaller than that of the secondary, because its temperature is much lower, we neglected both absorption and %%@
emission from the primary shock.

The parameter range for the model shock emission that allowed us to reproduce the observed light curve are presented in the %%@
second column of Table 1, the parameters of the best model are listed in the third column. The best parameters for the disk %%@
emission model represented by equation (1) are: $a=3.1\times 10^{-5}$, $b=-0.228$, $c=7.50$, $m= -0.004$ and $n = 0.94$, %%@
the time $t$ was measured from 29 June 2003 (JD = 2452819).
 
We found the best fitting  for  $e = 0.95$ and $\alpha = 4$, although solutions were also possible for $0.93 \leq e \leq %%@
0.95$, in which case $5.5 \geq \alpha \geq 4$. For the best model the orbital angle at periastron is $\Phi = -30^\circ \pm %%@
5^\circ$, while for the other eccentricities it varied  in the interval $-50^\circ \leq \Phi \leq -30^\circ$. For such high %%@
eccentricities the differences in the epoch of periastron passage and conjunction relative to the maximum in the light %%@
curve (June 29, defined as $t=0$ in our graphs) are very small: $3.2 \geq t(\phi = 0) \geq 1.5$ and $-1 \leq t(\phi = \Phi) %%@
\leq 0.1$ respectively, where $t$ is measured in days.

   \begin{figure}
    \centering
   \includegraphics[bb=95 200 390 705,width=8cm]{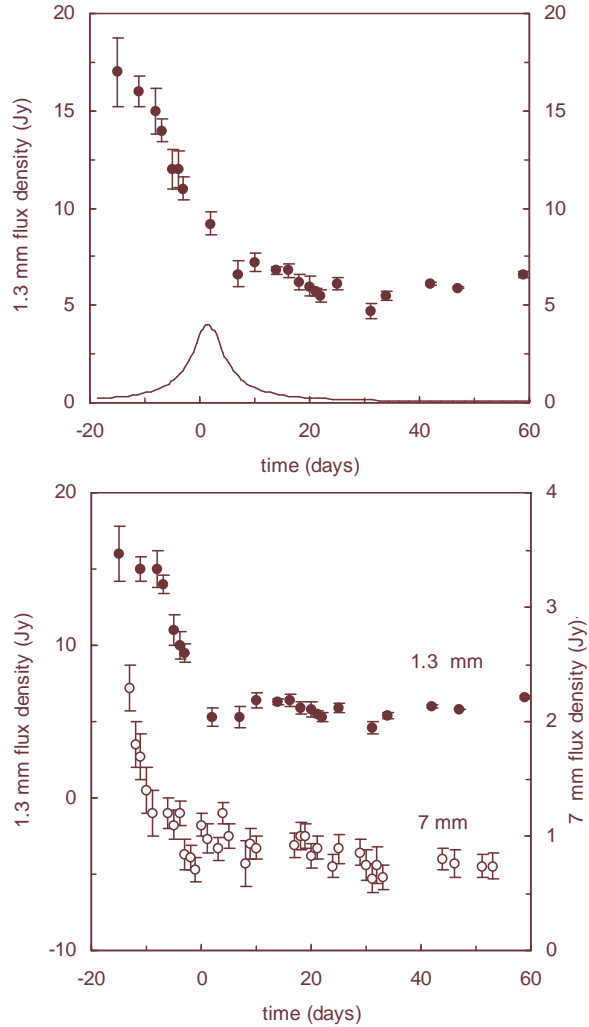}
      \caption 
	  {Model for the 1.3 mm light curve. Upper panel: 1.3 mm data (filled circles) and  model shock emission (solid line), %%@
obtained with the parameters presented in Table 1; lower panel: expected disk emission, both at 1.3 and 7 mm, calculated as %%@
the difference between the observations and the shock model flux density. Zero in the time axis corresponds to June 29, %%@
2003 (JD = 2452819).}
         \label{figure5}
   \end{figure}

The comparison of the model with the observations can be seen in Figures 4. The upper panel  shows the total fitted model %%@
emission (disk plus shock), together with the observed 7 mm flux density (circles) and the model disk  emission. The lower %%@
panel shows the shock emission model (continuous line) and the observed shock contribution, calculated as  the difference %%@
between the total observed flux density and the disk model contribution. This panel also shows the residuals, calculated as %%@
the difference between the observed flux density and the added disk and shock model contribution. 

Although the 1.3 mm light curve has not enough temporal resolution to show a peak in the light curve, we know that the %%@
predicted emission must be proportional to the optical depth at each point in the orbit since, as discussed before, the %%@
region should  be  optically thin at this wavelengths. 
As in the 7 mm model, we assumed that the optical depth varies as $[d_{\rm s,p}(\phi)/d_{\rm s,p}(0)]^{-\alpha}$, its  %%@
value at periastron can be adjusted to give the right flux density and  $\alpha$ is already determined by the 7 mm emission %%@
model. The  model emission at 1.3 mm for $\alpha = 4$ can be seen in the upper panel of Figure 5, together with the %%@
observations.The lower panel shows the expected  disk emission, both at 1.3 and 7 mm, calculated as the difference between %%@
the observations and the shock model flux density. 
   
\begin{table}
%\centering
%\begin{minipage}{120mm}
 \caption{Parameters for the $\eta$ Carinae binary system derived from mm-wave shock emission}
 \begin{tabular}{l@{}c@{}c@{}}
 \hline
 Parameter & Interval   &   Best value \\
\hline
$\beta$ (degrees)                     & ~~~~~~$40 - 60$  &  ~~~~~~56~~~~~~  \\
$e$                          &  ~~~~~~$0.92 - 0.95$   & ~~~~~~0.95~~~~~~\\
$t_p$ (since June 29)        &   ~~~~~~$1.5 - 3.2$ &  ~~~~~~1.5~~~~~~\\
$t_c$ (since June 29)        &  ~~~~~~$-1 - 0.1$    & ~~~~~~0.1~~~~~~\\
$\Phi$ (degrees)           & ~~~~~~$(-30) - (-50)$ &   ~~~~~~-30~~~~~~ \\
\hline
\end{tabular}
\end{table}

\begin{figure}
    \centering
   \includegraphics[bb=90 465 360 700,width=8cm]{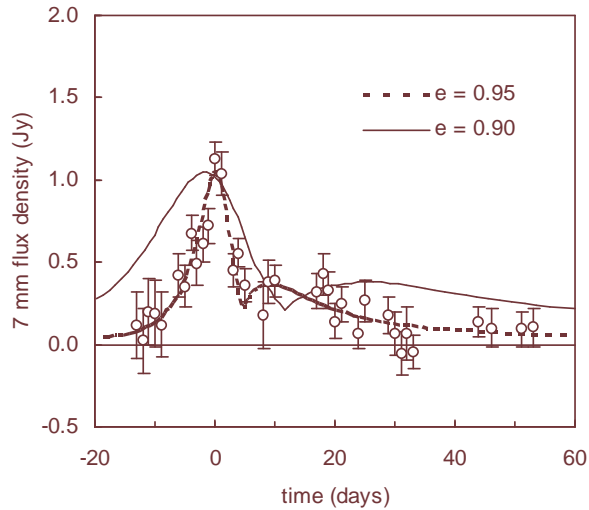}
      \caption 
	  {Model flux densities for eccentricities 0.9 (light line) and 0.95 (heavy line), together with the observed flux %%@
density (circles). Zero in the time axis corresponds to June 29, 2003 (JD = 2452819).}
         \label{figure6}
   \end{figure}

\section{Discussion}

The orbital parameters of the $\eta$ Carinae binary system were estimated under the assumption that the mm-wave emission %%@
during the low excitation phase is produced both by an ionized disk surrounding the system and by the shock region formed %%@
by wind-wind collision. 

In our model for the shock emission the eccentricity is the best defined parameter; small differences in its value cause a %%@
large broadening of the radio light curve. As mentioned in the last section, fitting to the light curve were obtained for %%@
$0.93 \leq e \leq 0.95$, these values are  slightly  higher than the value of 0.9 found by Pittard \& Corcoran (2002) in %%@
their numerical simulations of the X-ray spectral behavior; however this small difference produces large changes in our  7 %%@
mm shock emission model, as can be seen in Figure 6. These differences cannot be compensated by a decrease in the cone %%@
aperture angle $\beta$ or an increase in the dependence of the optical depth with the distance between the shock and the %%@
star.  The aperture angle  of the cone $\beta$, which represents the shock surface asymptotic behavior, was varied between %%@
$40^\circ$ and $60^\circ$. These values correspond to $0.1 < \eta < 0.2$, similar to those found by Pittard \& Corcoran  %%@
(2002). Due to the high eccentricity, the epochs of periastron passage and opposition  are very close, differing in only a %%@
few days, even though the phase for opposition is not small $(-30^\circ < \Phi< -50^\circ)$.

The fitting of the light curve described in the last  section was made under the assumption that periastron occurs close to %%@
conjunction, that is, when the secondary star passes between the observer and $\eta$ Carinae. If periastron occurs close to %%@
opposition, the role of primary and secondary shock emission are reversed. The behavior of the light curve for $\;\mid \phi %%@
+ \Phi \mid\; \leq \beta $ would be similar, but for larger phase angles the contribution of the secondary shock emission, %%@
much higher than that of the primary, would produce an increase in the model light curve, incompatible with the %%@
observations. We consider this result  as additional evidence for the case presented in Paper I, in which we explained the %%@
X-ray light curve and shell-like behavior of the $\eta$ Carinae binary system by assuming the secondary star positioned %%@
between $\eta$ Carinae and the observer close to periastron passage.

The  eccentricity values found in our model are compatible with the variation  of the H recombination lines radial %%@
velocities with orbital phase, as measured by Damineli et al. (1997, 2000) and Davidson (1997), if we assume that the lines %%@
are produced in the primary shock instead of in the stellar  atmosphere (Hill, Moffat \& St-Louis 2000, 2002). With these %%@
large values for the eccentricity, the distance from the shock to $\eta$ Carinae at periastron turns out to be less than 1 %%@
AU and, depending on the extension of its atmosphere, it can change the shock behavior on the primary side. However, we do %%@
not expect too much changes in our emission model, which depends mainly on the secondary shock. 

Finally, the linear extension $R$ of the shock cone can be calculate from the actual value of the observed 7 mm flux %%@
density, resulting in $R = 4\times 10^{14}$ cm for a plasma temperature of $10^7$ K. 
This value is valid at the orbital phase corresponding to the peak of the 7 mm emission. 
If this value were not constant along the orbit, the width of the peak would be different. In the analytical treatment of %%@
the shock properties, Usov (1991) found that the size increases with the distance between the stars, which will broadened  %%@
the  peak model. However, for $\;\mid \phi + \Phi \mid\; <  \beta $  the difference in distance is less than a factor of %%@
two 

\section{Conclusions}

In this paper we used the 1.3 and 7 mm light curves of $\eta$ Carinae during the 2003.5  low excitation phase to determine %%@
the orbital parameters of the binary system. We assumed that the emission is produced both in the extended disk surrounding %%@
the $\eta$ Carinae binary system, as proposed by Duncan \& White (2003), and at the shock region, the later responsible for %%@
the sharp peak seen in the 7 mm light curve during the decreasing phase of the disk emission. 
We calculated the form of the momentum equilibrium surface and the physical conditions in the material behind the primary %%@
and secondary shocks and showed that the emission could be produced in the non distorted part of the asymptotic cone formed %%@
by the two shocks, as proposed by Abraham et al. (2005). 
From the ratio of the 7 mm and 1.3 mm observed flux densities, we concluded that the emitting region should be optically %%@
thick at 7 mm and optically thin at 1.3 mm. For this reason, the shape of the 7 mm light curve close to periastron was %%@
completely determined by the value of the solid angle subtended by the emitting surface, which changes with the orbital %%@
position angle of the secondary star, and therefore, depends only on the orbital parameters. We found satisfactory %%@
agreement between models and observations for orbits with eccentricities $0.93 \leq e \leq 0.95$, orbital angle at %%@
periastron $-50^\circ \leq \Phi \leq -30^\circ$ and cone aperture angle  $40^\circ < \beta < 60^\circ$, compatible with %%@
what was expected from other observations, like X-ray fluxes and radial velocities of the H recombination lines (Pittard \& %%@
Corcoran 2002, Damineli et al. 2000, Davidson et al. 2000).  

The epoch of periastron passage and opposition varied between  $3.2 \geq t(\phi = 0) \geq 1.5$ and $-1 \leq t(\phi = \Phi) %%@
\leq 0.1$ respectively, where $t$ is measured in days from June 29, 2003.

Finally we must mention that the model presented here is a very simplified version of a very complex process and that %%@
confirmation of the orbital parameters presented here should be obtained from detailed numerical simulations.

\section*{Acknowledgments}

      This work was supported by the Brazilian Agencies FAPESP, CNPq and FINEP.


\begin{thebibliography}{99}
%\bibitem[\protect\citeauthoryear{Abraham \& Kokubun}{1992}]{abraham92} Abraham Z., Kokubun F., 1992,
 %A\&A, 257, 831
%\bibitem[\protect\citeauthoryear{Abraham \& Damineli} {1997}]{abr97} Abraham Z., Damineli A., 1997, in Luminous Blue
% Variables: Massive Stars in Transition, ASP Conference Series 120, p 294
\bibitem[\protect\citeauthoryear{Abraham \& Damineli}{1999}]{abraham99} Abraham Z., Damineli A., 1999, in Carinae at the %%@
Millenium, ASP Conference Series 179, San Francisco, p263 
\bibitem[\protect\citeauthoryear{Abraham et al.}{2002}]{abraham02} Abraham Z., Damineli A.,
 Durouchoux, P., Nyman
 L-\AA., McAuliffe F., 2002, in Cosmic Masers: from Proto-stars to Black Holes, ASP Conference
 Series 206, San Francisco, p234
\bibitem[\protect\citeauthoryear{Abraham et al.} {2005}]{abr05} Abraham, Z., Falceta-Gon\c calves,
 D., Dominici, T. P. et al. 2005, A\&A, 437, 977 
 \bibitem[\protect\citeauthoryear{Churchwell et al.} {1992}]{church92} Churchwell, E., Bieging, J. H., van der Hucht, K. A. %%@
et al., 1992, ApJ, 393, 329
%\bibitem[{Corcoran et al.} {2001a}]{corcoran01} Corcoran M. F., Ishibashi K., Swank J. H., Petre,
 %R., 2001a, ApJ, 547, 1034
%\bibitem[\protect\citeauthoryear{Corcoran et al.} {2001b}]{corcoran01b} Corcoran M. F., Swank J. H., Petre R., et al.,
% 2001b, ApJ, 562, 1031
\bibitem[\protect\citeauthoryear{Corcoran} {2005}]{corcoran05} Corcoran, M. F. 2005, AJ, 129, 2018 
\bibitem[\protect\citeauthoryear{Cox et al.} {1995a}]{cox95} Cox P., Mezger P. G., Sievers A. et
 al., 1995, A\&A, 297, 168 
%\bibitem[\protect\citeauthoryear{Cox et al.} {1995b}]{cox95b} Cox, P., Martin-Pintado, J.,
% Bachiller, R. et. al., 1995b, A\&A, 295, L39
%\bibitem[\protect\citeauthoryear{Cox} {1999}]{cox99} Cox, P., 1999 in Eta Carinae at the Millenium, ASP Conference Series %%@
179,
% San Francisco, p277
%\bibitem[\protect\citeauthoryear{Damineli} {1996}]{damineli96} Damineli, A.,  1996, ApJ, 460, L49
\bibitem[\protect\citeauthoryear{Damineli et al.} {1997}]{damineli97} Damineli, A. \& Conti, P., Lopes, D., 1997,
 NewA, 2, 107
\bibitem[\protect\citeauthoryear{Damineli et al.} {2000}]{damineli00} Damineli, A., Kaufer, A., Wolf, B., Stahl, O., Lopes, %%@
D. \& Ara\'{u}jo, F. 2000, ApJ, 528, L101
\bibitem[\protect\citeauthoryear{Davidson}  {1997}]{davidson97} Davidson K., 1997, NewA, 2, 397  
%\bibitem[\protect\citeauthoryear{Davidson \& Hunphreys}{1997}]{dav97}Davidson, K., %Humphreys, R. 1997, Ann. Rev. Astron. %%@
Astrophys., 35, 1
\bibitem[\protect\citeauthoryear{Davidson et al.} {2000}]{davidson00} Davidson, K., Ishibashi, K., Gull, T. R.,
 Humphreys, R., Smith, N., 2000, ApJ, 530, L107 
%\bibitem[\protect\citeauthoryear{Dougherty et al.} {1996}]{dough96} Dougherty, S. M., Williams, P. M., van der Hucht, K. %%@
A., %Bode, M. F., Davis, R. J. 1996, MNRAS, 280, 963
%\bibitem[\protect\citeauthoryear{Dougherty \& Williams} {2000}]{dough00} Dougherty, S. M., Williams, P. M., 2000, MNRAS, %%@
319, %1005
%\bibitem[\protect\citeauthoryear{Dougherty et al.} {2003}]{dough03} Dougherty, S. M., Pittard, J. M., Kasian, L. et al. %%@
2003, %A\&A, 409, 217
\bibitem[\protect\citeauthoryear{Duncan et al.} {1997}]{duncan97} Duncan, R. A., White, S. M., Lim,
 J., 1997, MNRAS, 290, 680 
\bibitem[\protect\citeauthoryear{Duncan \& White} {2003}]{duncan03} Duncan, R. A., White, S. M. 2003, MNRAS, 338, 425 
\bibitem[\protect\citeauthoryear{Eichler \& Usov}{1993}]{eichler93} Eichler, D., Usov, V. 1993, ApJ
402, 271
\bibitem[\protect\citeauthoryear{Falceta-Gon\c calves et al.} {2004}]{falceta04} 
Falceta-Gon\c calves D.,Jatenco-Pereira V., Abraham Z, 2005, (Paper I) MNRAS, 357, 895 
%\bibitem[\protect\citeauthoryear{Girard \& Willson} {1987}]{girard87} Girard, T., Willson, L. A., 1987, A\&A
%183, 247
\bibitem[\protect\citeauthoryear{Hill, Moffat \& St-Louis}{Hill et al.} {2000}]{hill00} Hill G., Moffat A., St-Louis A., %%@
2000, MNRAS, 318, 402
\bibitem[\protect\citeauthoryear{Hill, Moffat \& St-Louis}{Hill et al.} {2002}]{hill02} Hill G., Moffat A., St-Louis A., %%@
2002, MNRAS, 335, 1069
%\bibitem[\protect\citeauthoryear{Hillier et al.} {2001}]{hillier01} Hillier D., Davidson K., Ishibashi K., Gull T., 2001, %%@
ApJ, %553, 837
\bibitem[\protect\citeauthoryear{Ishibashi et al.} {1999}]{ishibashi99} Ishibashi K. Corcoran M., Davidson K., Swank J., %%@
Petre R., Drake S., Damineli A., White S., 1999, ApJ, 524, 983 
\bibitem[\protect\citeauthoryear{Laj\'us et al.} {2003}]{lajus03} Laj\'us, F. E., Gamen, R., Schwartz, M. et al. 2003, %%@
IBVS, 5477, 1
\bibitem[\protect\citeauthoryear{Martin et al.} {2004}]{martin04} Martin, J. C., Koppelman, M. D. et al. 2004, ApJ, 127, %%@
2352 
%\bibitem[\protect\citeauthoryear{Mathews \& Doane} {1990}]{mathews90} Mathews W., Doane J., 1990, ApJ, 352, 423
%\bibitem[\protect\citeauthoryear{Moran et al.} {1989}]{moran89} Moran, J. P., Davis, R. J., Spencer, R. E., Bode, M. F., %%@
%Taylor, A. R. 1989, Nature, 340, 449
%\bibitem[\protect\citeauthoryear{Niemela et al.} {1998}]{niemela98} Niemela, V. S., Shara, M. M., Wallace, D. J., Zurek, %%@
D. %R., Moffat, A. F. J., 1998, AJ, 115, 2047
\bibitem[\protect\citeauthoryear{Pittard et al.} {1998}]{pittard98} Pittard J. M., Stevens I. R., Corcoran M. F., Ishibashi %%@
K., 1998, MNRAS, 299, L5 
\bibitem[\protect\citeauthoryear{Pittard \& Corcoran} {2002}]{pittard02} Pittard J. M., Corcoran M. F., 2002, A\&A, 383, %%@
636 
%\bibitem[\protect\citeauthoryear{Retalack} {1983}]{ret83} Retalack, D. S. 1983, MNRAS 204, 669
%\bibitem[\protect\citeauthoryear{Seward et al.} {2001}]{seward01} Seward F. D., Butt Y. M., Karovska M., Prestwich A., %%@
%Schlegel E. M., 2001, ApJ, 553, 832
\bibitem[\protect\citeauthoryear{Smith et al.} {2004}]{Smith04} Smith, N., Morse, J. A., Collins, N.
 R., Gull, T. R. 2004, ApJ, 610, L105 
%\bibitem[\protect\citeauthoryear{Spitzer} {1978}]{spitzer78} Spitzer L., 1978, Physical processes in the interstellar %%@
medium, %New York, Wiley.
\bibitem[\protect\citeauthoryear{Stahl et al.} {2005}]{stahl05} Stahl O., Weis, K, Bomans, D. J. et al. 2005, A\&A, 435, %%@
303 
\bibitem[\protect\citeauthoryear{Steiner \& Damineli} {2004}]{steiner04} Steiner, J. E., Damineli, A. 2004, ApJ, 612, L133 
\bibitem[\protect\citeauthoryear{Stevens et al.} {1992}]{Stevens92} Stevens, I. R., Blondin, J. M., Pollock, A. M. T., %%@
1992, ApJ, 386, 265
\bibitem[\protect\citeauthoryear{Usov} {1991}]{uso91} Usov V. V.,1991, MNRAS, 252, 49
%\bibitem[\protect\citeauthoryear{Usov} {1992}]{usov92} Usov V. V.,1992, ApJ, 389, 635
\bibitem[\protect\citeauthoryear{van Genderen} {1999}]{vang99} van Genderen, A. M., Sterken, C., de
 Groot, M., Burki, G.  1999, A\&A, 343, 847 
\bibitem[\protect\citeauthoryear{van Genderen \& Sterken} {2003}]{vang03} van Genderen, A. M.,
 Sterken, C. 2003, A\&A, 423, L1
\bibitem[\protect\citeauthoryear{Weis et al.} {2005}]{weis05} Weis, K., Stahl, O., Bomans, D. J. et al. 2005, AJ, 129, 1694 
\bibitem[\protect\citeauthoryear{Whitelock  et al.} {2003}]{whi03} Whitelock, P. A., Marang, F., Crause, L, Corcoran, M., %%@
2003 IAUC, 8160
\bibitem[\protect\citeauthoryear{Whitelock  et al.} {2004}]{whi04} Whitelock, P. A.,Feast, M. F.,  Marang, F.,  Breedt, E.
 2004, 
MNRAS, 352, 447 
%\bibitem[\protect\citeauthoryear{Williams et al.} {1997}]{will97} Williams, P. M., Dougherty, S. M., Davis, R. J. et al., %%@
%1997, MNRAS, 289, 10

\end{thebibliography}
\end{document}